\documentclass[aps,prb,twocolumn,superscriptaddress]{revtex4-1}
\usepackage{graphicx}
 \usepackage{color}
\usepackage{amsfonts}
\usepackage{isomath}
\usepackage{amsmath}
\usepackage{amsthm}
\usepackage{amsfonts}
\usepackage{braket}

\usepackage{yhmath}

\begin{document}

\title{Resonantly enhanced photoemission from topological surface states in MnBi$_6$Te$_{10}$}

\author{Paulina Majchrzak}
\affiliation{Department of Applied Physics, Stanford University, Stanford, CA, USA}
\author{Alfred J.  H.  Jones}
\affiliation{Department of Physics and Astronomy, Interdisciplinary Nanoscience Center, Aarhus University, 8000 Aarhus C, Denmark}
\author{Klara Volckaert}
\affiliation{Department of Physics and Astronomy, Interdisciplinary Nanoscience Center, Aarhus University, 8000 Aarhus C, Denmark}
\author{Xing-Chen Pan}
\affiliation{Advanced Institute for Materials Research, Tohoku University, Sendai 980-8577, Japan}
\author{Philip Hofmann}
\affiliation{Department of Physics and Astronomy, Interdisciplinary Nanoscience Center, Aarhus University, 8000 Aarhus C, Denmark}
\author{Yong P.  Chen}
\affiliation{Department of Physics and Astronomy, Interdisciplinary Nanoscience Center, Aarhus University, 8000 Aarhus C, Denmark}
\affiliation{Department of Physics and Astronomy and School of Electrical and Computer Engineering and Purdue Quantum Science and Engineering Institute, Purdue University, West Lafayette, IN 47907, USA}
\author{Jill A. Miwa}
\author{S{\o}ren Ulstrup}
\email{ulstrup@phys.au.dk}
\affiliation{Department of Physics and Astronomy, Interdisciplinary Nanoscience Center, Aarhus University, 8000 Aarhus C, Denmark}

\begin{abstract}
The dispersion of topological surface bands in MnBi$_2$Te$_4$-based magnetic topological insulator heterostructures is strongly affected  by band hybridization and is spatially inhomogeneous due to varying surface layer terminations  on microscopic length scales.  Here,  we apply micro-focused angle-resolved photoemission spectroscopy with tunable photon energy from 18 to 30 eV to distinguish bulk valence and conduction bands from surface bands on the three surface terminations of MnBi$_6$Te$_{10}$. We observe a strong enhancement of photoemission intensity from the topological surface bands at the Bi $O_4$ absorption edge, which is exploited to visualize a gapless Dirac cone on the MnBi$_2$Te$_4$-terminated surface and varying degrees of hybridization effects in the surface bands on the two distinct Bi$_2$Te$_3$-terminated surfaces.  \\ 

\end{abstract}

\maketitle

\section{Introduction}

Magnetic topological insulators support time-reversal symmetry broken topological states that give rise to a quantum anomalous Hall effect and dissipationless chiral edge states \cite{Yujun2020,Deng2021}. One of the most intensely studied materials in this class is MnBi$_2$Te$_4$ (MBT).  The non-trivial topology of MBT derives from the band inversion between the Te~5p$_z$ valence band and the Bi~6p$_z$ conduction band, while the intrinsic magnetism arises due to a finite magnetic moment associated with the Mn atoms \cite{otrokov2019prediction,Vidal:2019,Klimovskikh2020}.  Indeed,  in the family of MnBi$_{2+2n}$Te$_{4+3n}$ ($n=0,1,2,\ldots$) heterostructures where MnBi$_2$Te$_4$ layers are interspaced by $n$ layers of Bi$_2$Te$_3$,  several quantum anomalous Hall regimes \cite{Yujun2020,Deng2021,Wang:2023,Gao:2023,Li:2024} have been observed along with axion insulator states \cite{Liu:2020,Hyun:2020} and interesting ferromagnetic phases \cite{Xiaolong:2022a,Chenhui:2022c,Tcakaev:2023}.  In these materials, a septuple layer (SL) of MnBi$_2$Te$_4$ adjacent to a quintuple layer (QL) of Bi$_2$Te$_3$ interacts via the van der Waals force, which makes exfoliation of thin flakes with thickness-tunable magnetism possible \cite{hu2020van}. 

The salient electronic and magnetic properties of MnBi$_{2+2n}$Te$_{4+3n}$ derive from the surface electronic band structure. In particular,  hybridization effects in the surface bands, the gap size between the bulk valence band (VB) and conduction band (CB), the doping concentration and the size of the exchange gap in the topological surface state (TSS) dictate whether quantum anomalous Hall regimes can be attained in electron transport devices \cite{otrokov2019prediction,Jiaheng2019,wu2019natural,zhang2019topological,Vidal:2019,Vidal2:2019,Hao:2019,Li2019,Chen2019,Runzhe:2024}.  Angle-resolved photoemission spectroscopy (ARPES) has revealed a tendency for this family of materials to be strongly electron-doped with a complex interplay of surface and bulk VB and CB states \cite{Zhicheng2022,Runzhe:2024}. The nature of the TSS and whether it displays a gap have been strongly debated \cite{Shikin:2021,Zhicheng2022,Hengxin:2022,Runzhe:2024,Vyazovskaya:2025}.  Micro-focused ARPES (microARPES) measurements using vacuum ultraviolet (VUV) lasers operating with photon energies in the range 6-7~eV have been instrumental in resolving the detailed TSS dispersion for a given surface termination of the MnBi$_{2+2n}$Te$_{4+3n}$ compounds due to high energy- and momentum-resolution \cite{Xiao:2020,Hu2020,Tian:2020,Vidal2021}.  More recently, spin-resolved ARPES enabled description of the surface state wavefunction in the $n=(0,1)$ compounds,  revealing information on its spin and orbital texture \cite{Han:2025}. However, the lack of photon energy tunability with VUV sources precludes measuring bulk band dispersion, which complicates the study of surface and bulk band interactions and how these affect the dispersion of the TSS.

Here, we determine the electronic structure of the $n=2$ compound, MnBi$_6$Te$_{10}$, over a wide photon energy range, which permits tuning of the out-of-plane momentum, using micro-scale synchrotron radiation focused with an achromatic capillary optic. This procedure makes it possible to disentangle surface states from bulk states as only the latter are dispersive with out-of-plane momentum. Furthermore,  the elemental character of the bands gives rise to photon energy-dependent photoemission matrix element effects,  which we exploit to map the detailed topological surface state dispersion on all terminations of MnBi$_6$Te$_{10}$.

\begin{figure*}[t!] 
\begin{center}
\includegraphics[width=1.0\textwidth]{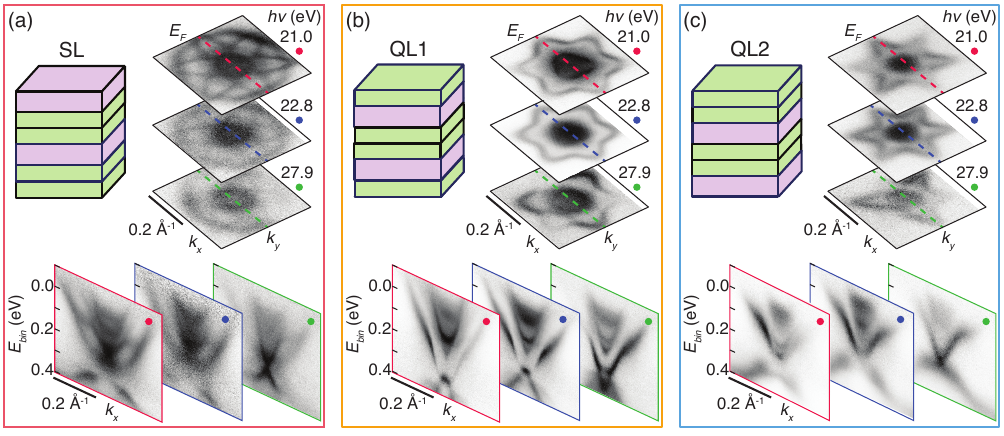}
\caption{(a)-(c) $(k_x,k_y)$- and $(k_x,E_{bin})$-dependent ARPES intensity for (a) SL,  (b) QL1 and (c) QL2 terminations of MnBi$_6$Te$_{10}$ at photon energies of 21.0~eV (red dot),  22.8~eV (blue dot) and 27.9~eV (green dot).  Definitions of the structures are given in the insets with septuple layers drawn as pink boxes and quintuple layers drawn as green boxes.  The $(k_x,k_y)$-dependent spectra are constant energy surfaces extracted at the Fermi energy, $E_F$.  $(k_x,E_{bin})$-cuts are extracted along the dashed lines indicated on the constant energy surfaces.}
\label{fig:1}
\end{center}
\end{figure*}

\section{Experimental}
Bulk crystals of MnBi$_6$Te$_{10}$ were prepared using the flux method with Mn:Bi:Te powder mixing ratio of 1:10:16 in sealed quartz tubes. The enclosed mixture was heated to 900~$^{\circ}$C and cooled to 590~$^{\circ}$C,  followed by flux removal via centrifuge upon cooling down.

The microARPES measurements were performed at the SGM4 beamline of the ASTRID2 light source at Aarhus University \cite{Jones:2025}.  MnBi$_6$Te$_{10}$ crystals were cleaved via epoxy-glued top posts in a pressure better than 10$^{-8}$~mbar using the load lock of the microARPES system. The synchrotron beam was focused to a spot size with lateral dimensions of 4 $\mu$m using a capillary optic,  which made it possible to locate optimally cleaved domains of a distinct surface termination of MnBi$_6$Te$_{10}$.  The smallest domains observed were on the order of $(20 \times 20)$ $\mu$m$^2$, which were estimated from real space maps of the photoemission intensity obtained by raster scanning the sample under the beam in 5 $\mu$m steps as described in detail in Ref.~\citenum{Jones:2025}. 

Measurements were performed at room temperature.  The total energy resolution combining beamline and electron analyser settings amounted to 21~meV, and the momentum resolution was 0.01~\AA${^{-1}}$.

\section{Results and Discussion}

\begin{figure*}[t!] 
\begin{center}
\includegraphics[width=1.0\textwidth]{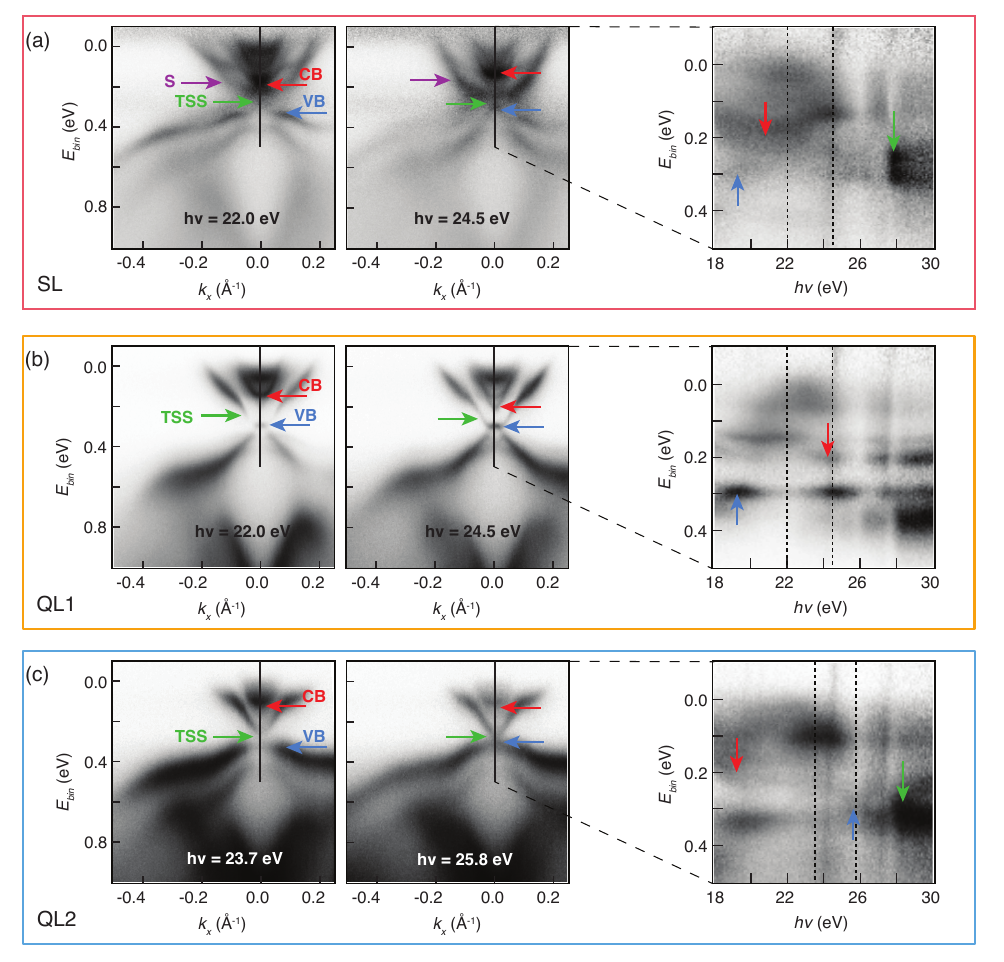}
\caption{(a)-(c) Photon energy tuning of ARPES intensity around normal emission for (a) SL,  (b) QL1 and (c) QL2 terminations of MnBi$_6$Te$_{10}$.  Arrows demarcate bulk conduction band (CB) and valence band (VB) features,  as well as surface bands (S) and  topological surface state (TSS) bands.  The $(k_x,E_{bin})$-dependent intensity has been extracted at the stated photon energies,  which are demarcated by vertical dotted lines on the shown $(hv,E_{bin})$-cuts at normal emission.  The vertical bars on the $(k_x,E_{bin})$-spectra at normal emission ($k_x = 0$) demarcate the EDCs measured as a function of $hv$. } 
\label{fig:2}
\end{center}
\end{figure*}

Cleaving MnBi$_6$Te$_{10}$ exposes three types of surface terminations with the ordering of SL and QL units sketched in Figs.~\ref{fig:1}(a)-(c).  The measured surface electronic structure is presented for each termination via the photoemission intensity at the Fermi level at photon energies of 21.0, 22.8 and 27.9~eV and corresponding $E(k_x)$-dispersion cuts through the $\bar{\mathrm{M}}-\bar{\mathrm{\Gamma}}-\bar{\mathrm{M}}$ high-symmetry direction of the surface Brillouin zone.  The system is observed to be strongly $n$-doped, which is consistent with other reports of the pristine MBT systems \cite{Xiao:2020,Hu2020,Tian:2020,Vidal2021}. SL-terminated MnBi$_6$Te$_{10}$ is characterized by a six-petal flower-shaped Fermi contour with an inner circle that stems from bulk CB states and an outer circle that is commonly ascribed to a Rashba-like state driven by the presence of surface impurities \cite{Yan2021, Volckaert:2023},  as seen in Fig.~\ref{fig:1}(a). For the surface termination where a single QL layer is situated above a SL layer, denoted as QL1, the Fermi contour instead resembles a warped hexagon in addition to the bulk CB inner circle, as shown in Fig.~\ref{fig:1}(b).  Finally,  in the case where the surface is terminated by a QL bilayer, labelled as QL2,  the Fermi contour displays a star-like shape in addition to the central bulk states, as seen in Fig.~\ref{fig:1}(c). The corresponding dispersion cuts reveal distinct surface bands in addition to parabolic bulk CB states with photon energy-dependent intensity variations within the bands, which result from varying degrees of surface band hybridization effects on the three terminations, as we discuss in the following.

\begin{figure*}[t!] 
\begin{center}
\includegraphics[width=1\textwidth]{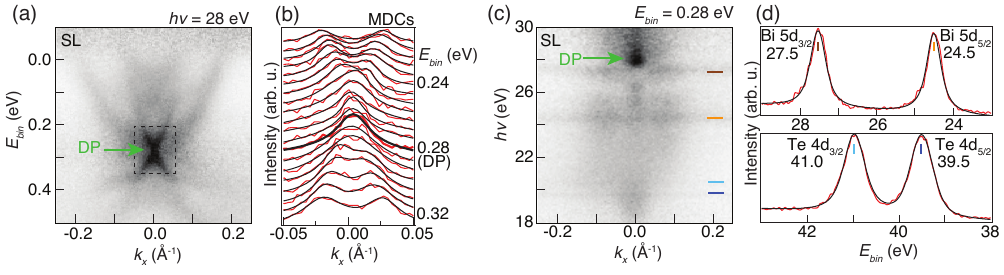}
\caption{(a) ARPES dispersion on the SL termination of MnBi$_6$Te$_{10}$ at a photon energy of 28~eV.  The Dirac point (DP) of the topological surface state is indicated by a green arrow.  (b) Stack of MDCs (red curves) extracted from the ARPES intensity outlined by a dashed box in (a) with double Lorentzian fits (black curves).  The MDC and fit curves at the Dirac point are highlighted by increased thickness.  (c) $(k_x,hv)$-dependent ARPES intensity at the binding energy of the Dirac point.  (d) Core level spectra (red curves) of the Bi 5d and Te 4d binding energy regions with Lorentzian peak fits (black curves).  Fitted peak positions are stated in units of eV and indicated by ticks.  Correspondingly colored ticks demarcate intensity from the same core levels in (c).  Core level measurements were taken with a photon energy of 140~eV.}
\label{fig:3}
\end{center}
\end{figure*}

A more detailed understanding of the underlying surface or bulk character of the MnBi$_6$Te$_{10}$ bands can be obtained by varying the photon energy of the synchrotron probe beam, as this changes the out-of-plane momentum of photoelectrons. This approach is applied to the SL, QL1 and QL2 terminations as shown in Fig.~\ref{fig:2}.  In each case, two $E(k_x)$-dispersion cuts are shown down to a binding energy of 1.0~eV.  Energy distribution curves (EDCs) at normal emission are presented down to a binding energy of 0.5~eV as a function of photon energy from 18 to 30~eV.  The photoemission intensity has been divided by the Fermi-Dirac distribution plus a constant intensity offset in order to visualize the dispersion around $E_F$.  Bulk VB and CB features are identified and shown via blue and red arrows, respectively. The bulk VB maximum is located at a binding energy of $\mathbin{{\approx}{0.3}}$~eV and the bulk conduction band minimum at $\mathbin{{\approx}{0.2}}$~eV, which leaves a gap of $\mathbin{{\approx}{0.1}}$~eV for the genuine surface states at normal emission as highlighted by green arrows in Fig. ~\ref{fig:2}. Note that the aforementioned surface impurity-driven Rashba-like state on the SL termination is demarcated by a purple arrow in Fig.~\ref{fig:2}(a). The presence of this state clearly distinguishes the SL surface terminations from the QL1 and QL2 terminations.

\begin{figure}[t!] 
\begin{center}
\includegraphics[width=0.5\textwidth]{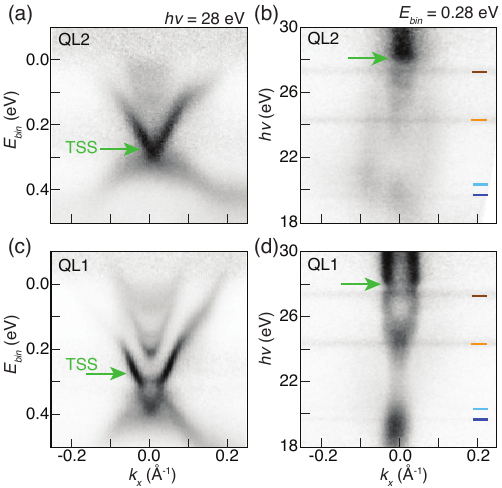}
\caption{(a)-(b) ARPES intensity of the QL2 termination of MnBi$_6$Te$_{10}$ presented via (a) $(k_x,E_{bin})$-cut at a photon energy of 28~eV and (b) $(k_x,hv)$-cut at a binding energy of 0.28~eV.  (c)-(d) Corresponding ARPES data for the QL1 termination of MnBi$_6$Te$_{10}$.  Topological surface bands are indicated by green arrows.  Intensity from Bi 5d and Te 4d core levels is demarcated by ticks following the analysis in Fig.~\ref{fig:3}(c).}
\label{fig:4}
\end{center}
\end{figure}

The relatively flat binding energy dispersion with photon energy is to be expected, because the out-of-plane lattice constant is 102 \AA~\cite{ALIEV2019443},  such that the measured photon energy range corresponds to an out-of-plane momentum range crossing around 10 BZs.  The discernible smooth variations are therefore mainly caused by photoemission matrix element effects that reflect the orbital character of the initial state wavefunction.
Around a photon energy of 28~eV, we observe a strong enhancement of photoemission intensity from bands that are located in the bulk band gap, as highlighted by the green arrows in the photon energy-dependent EDCs in Fig.  \ref{fig:2}. We explore this effect in more detail for the SL termination in Fig.~\ref{fig:3} and the QL terminations in Fig.~\ref{fig:4}.

Figure \ref{fig:3}(a) presents an $E(k_x)$ ARPES cut along the $\bar{\mathrm{M}}-\bar{\mathrm{\Gamma}}-\bar{\mathrm{M}}$ direction for the SL termination at a photon energy of 28~eV.  Intriguingly, the TSS surface state in the bulk band gap is strongly enhanced, which permits a detailed extraction of the Dirac point (DP) energy via momentum distribution curve (MDC) fits, as shown in Fig.~\ref{fig:3}(b). The DP binding energy is determined to be 0.28~eV.  Moreover, the dispersion displays minimal interaction with the bulk VB and CB states in the immediate vicinity of the bulk band edges.  

The intensity enhancement of the DP crossing is explored in further detail in Fig.~\ref{fig:3}(c),  which presents a photon energy-dependent scan of the MDC extracted at the DP binding energy. In addition to the sharp DP peak around normal emission and broad features from the CB at higher momenta, horizontal streaks of intensity are visible, which are marked by brown and orange ticks.  These features are Bi 5d$_{3/2}$ and 5d$_{5/2}$ core levels, which are observed at the given binding energy because the undulator insertion device in the synchrotron produces relatively intense second order light.  As the photon energy of the second order radiation becomes equal to or larger than twice the core level binding energy, a faint signal from these core levels is juxtaposed in the measured photoemission intensity around the Fermi energy.  Additional faint intensity is seen around a photon energy of 20~eV (see light and dark blue ticks), which stem from photoemission from the Te 4d core levels through higher-order light produced by the undulator.  Note that these core level features are also visible as nearly vertical streaks observable around the same photon energies in the photon energy scans in Fig \ref{fig:2}. The measured Bi 5d and Te 4d core level binding energies from the SL termination are presented in Fig.~\ref{fig:3}(d) for a photon energy of 140 eV.  The spin-orbit split components have nearly the same peak intensity, which appears inconsistent with their multiplicity.  However, the branching ratios of these peaks are strongly photon energy dependent due to the photon energy dependence of the photoemission cross sections for the spin-orbit split components and can approach unity under certain conditions \cite{Margaritondo:1979}.

Interestingly, the DP exhibits two step-like increases in the photoemission signal strength,  which coincide with the photon energy crossing the Bi $O_5$ and $O_4$ absorption edges, related to the electronic transition to the states around the Fermi level from Bi 5d$_{5/2}$ and Bi 5d$_{3/2}$ states, respectively.  In pure Bi, the $O_5$ and $O_4$ subshells are located at binding energies of 24.4 eV and 26.5 eV,  respectively,  which are determined via X-ray absorption spectroscopy in the extreme ultraviolet range \cite{BEARDEN:1967,Romain:2021}.  Note that the orbital-projected surface spectral density calculations in Ref.~\cite{Vidal2021} ascribe Bi and Te $J=1/2$ character to the TSS in MnBi$_4$Te$_7$. The enhancement of photoemission intensity is much more pronounced for the $\Delta J = \pm 1$ transition at the Bi $O_4$ absorption edge, which is consistent with the electrical dipole selection rules, and suggests a resonant photoemission process. We can exploit the resonance to visualize the TSS, otherwise only observed in the VUV 6-7~eV range in laser-based ARPES experiments.

Figure \ref{fig:4} presents the results of applying the resonant photoemission process to visualize the TSS on the QL1 and QL2 terminations. For the QL2 termination in Fig.~\ref{fig:4}(a), we observe the upper half of the linear dispersion becoming enhanced at the 28~eV photon energy and that the dispersion merges into the DP close to the bulk VB edge.  Fig.~\ref{fig:4}(b) presents the MDC at the DP as a function of photon energy,  exhibiting a similar behaviour as on the SL termination in Fig.~\ref{fig:3}(c).  On the QL1 termination, the situation is more complex as seen in Figs.~\ref{fig:4}(c)-(d). Two linear branches of the TSS are enhanced in the bulk gap, but they do not merge before reaching the bulk VB edge.  Instead, they appear to strongly hybridize with adjacent states, resulting in the abrupt splitting of the dispersion above a binding energy of 0.3 eV.  Two separate features are seen in Fig.~\ref{fig:4}(d) at the energy where we observe the DP on the other terminations.  These are the surface bands, which become increasingly more intense as the photon energy exceeds the binding energy of the Bi $O_5$ and $O_4$ edges.  At photon energies below these thresholds the surface band intensity is suppressed.  The separate feature at normal emission, which is enhanced around photon energies of 19 and 24 eV,  corresponds to the bulk VB edge. This switching behaviour of the intensity between bulk and surface bands around the bulk gap is also visible in the dispersion cuts in Fig.  \ref{fig:1}(b).

The complicated behaviours of the surface states on the three terminations have been ascribed to surface-bulk hybridization effects \cite{Xiao:2020},  or hybridization between surface layer-localized bands in the SL,  QL1 and QL2 layers \cite{Hengxin:2022}.  Fundamentally,  different terminations of surfaces can drastically change the shape and even position of the topological bands.  This has been shown by considering the  surface fermion parity for different terminations and surface orientations of BiSb \cite{Teo:2008,XieGang:2013,XieGang:2014}. The surface fermion parity is calculated with the implicit assumption that the surface termination goes through the same inversion centre that is used for the calculation of the bulk parity eigenvalues.  Changing the surface termination and moving it outside bulk inversion points would imply that the surface fermion parity must change as well,  which would be expected to have a significant impact on the dispersion and topology of the surface states in the MBTs.

In conclusion,  we have characterized the surface electronic structure of the three distinct surface terminations of MnBi$_6$Te$_{10}$.  We establish the bulk valence and conduction band edges at binding energies of 0.3 and 0.2~eV,  respectively, and measure the photoemission intensity dependence across a photon energy range of 18 to 30~eV.  Using a resonant photoemission process at the Bi $O_4$ edge, we determine a gapless Dirac point at a binding energy of 0.28~eV and a band dispersion without strong surface hybridization effects and minimal interactions with bulk states when crossing into the bulk band edges for the MnBi$_2$Te$_4$ septuple layer (SL) termination.  The bilayer Bi$_2$Te$_3$ quintuple-layer termination (QL2) displays a similar behavior,  although the linear dispersion is strongly diminished into the bulk valence bands.  On the single-layer Bi$_2$Te$_3$ quintuple-layer termination (QL1) we do not observe a Dirac crossing, but separate TSS bands with a splitting of the linear dispersion upon crossing the valence band edge.  Our results underscore the complexity of the TSS states on the terminations of MnBi$_6$Te$_{10}$,  which possibly originate from hybridization or termination-dependent topology effects.

\section{Data availability statement}
The data used in this study is available on Zenodo \cite{Zenodo}.

\section{Acknowledgements}
The work was funded/co-funded by the European Union (ERC grant EXCITE with project number 101124619). Views and opinions expressed are
however those of the author(s) only and do not necessarily reflect those of the European Union or the European
Research Council.  Neither the European Union nor the granting authority can be held responsible for them.  The authors acknowledge funding from the Novo Nordisk Foundation (Project Grant NNF22OC0079960),  VILLUM FONDEN under the Villum Investigator Program (Grant. No. 25931), the Danish Council for Independent Research (Grant Nos.  DFF-9064-00057B,  DFF-6108-00409 and 1026-00089B),  and the Aarhus University Research Foundation.

\end{document}